\documentclass[twocolumn,showpacs,preprintnumbers,amsmath,amssymb,
 aps,
 prb,
 lengthcheck
]{revtex4-1}
\usepackage{amssymb}
\usepackage{graphicx}% Include figure files
\usepackage{dcolumn}% Align table columns on decimal point
\usepackage{bm}% bold math
\usepackage{hyperref}% add hypertext capabilities
\usepackage{makeidx}
\usepackage{bbm}

\usepackage[english]{babel}
\usepackage[latin1]{inputenc}
\usepackage{color}
\usepackage{graphicx}
\usepackage{amsmath}

\usepackage{latexsym}
\usepackage{amsthm}

\makeindex
\def\>{\right\rangle}
\def\<{\left\langle}
\def\be{\begin{equation}}
\def\ee{\end{equation}}
\def\ba{\begin{array}{l}}
\def\ea{\end{array}}

\def\beq{\begin{eqnarray}}
\def\eeq{\end{eqnarray}}

\begin{document}

\preprint{APS/123-QED}

\title{Coulomb blockade microscopy of spin density oscillations and fractional charge in quantum spin Hall dots}

\author{G. Dolcetto$^{1,2,3}$, N. Traverso Ziani$^{1,2}$, M. Biggio$^{4}$, F. Cavaliere$^{1,2}$, M. Sassetti$^{1, 2}$}
 \affiliation{$^1$ Dipartimento di Fisica, Universit\` a di Genova, Via Dodecaneso 33, 16146, Genova, Italy.\\
$^2$ CNR-SPIN, Via Dodecaneso 33, 16146, Genova, Italy.\\
$^3$ INFN, Via Dodecaneso 33, 16146, Genova, Italy.\\
$^4$ Dipartimento di Ingegneria Navale, Elettrica, Elettronica e delle Telecomunicazioni, Universit\` a di Genova, Via Opera Pia 11a, 16145, Genova, Italy.}
\date{\today}

\begin{abstract}
\noindent
We evaluate the spin density oscillations arising in quantum spin Hall quantum dots created via two localized magnetic barriers. The combined presence of magnetic barriers and spin-momentum locking, the hallmark of topological insulators, leads to peculiar phenomena: a half-integer charge is trapped in the dot for antiparallel magnetization of the barriers, and oscillations appear in the in-plane spin density, which are enhanced in the presence of electron interactions.
Furthermore, we show that the number of these oscillations is determined by the number of particles inside the dot, so that the presence or the absence of the fractional charge can be deduced from the in-plane spin density.
We show that when the dot is coupled with a magnetized tip, the spatial shift induced in the chemical potential allows to probe these peculiar features.

\end{abstract}

\maketitle

\section{Introduction}\label{Intro}
Shortly after their theoretical prediction \cite{Bernevig06b} and experimental realization in HgTe quantum wells (QWs), \cite{Konig07, Konig08} two-dimensional (2D) topological insulators (TIs) \cite{Hasan10, Qi11} are still one of the most studied topics in condensed matter physics.
The fascination aroused by these systems comes both from the need of a deeper theoretical understanding and from the ambition of devising promising architectures for spintronics applications. \cite{Roth09}
Two dimensional topological insulators realize the quantum spin Hall (QSH) phase, \cite{Kane05a, Kane05b} characterized by an insulating bulk and metallic edge states, whose main property is spin-momentum locking. \cite{Wu06} In the presence of time-reversal (TR) symmetry, these helical edge states are topologically protected from backscattering. \cite{Bernevig06b}
The Fermi liquid paradigm fails in describing electrons in one dimension: \cite{Giamarchi03} QSH edges realize a new strongly correlated electron state, the helical Luttinger liquid (HLL). \cite{Wu06}
Despite TR symmetry is in general required to observe the conductance quantization, \cite{Bernevig06b,Konig07} lots of information about the QSH phase can be gained by breaking TR. Indeed, the presence of magnetic perturbations \cite{Meng09, Meng12, Kharitonov12, Qi08, Soori12, Timm12, Vayrynen11} or tunneling regions, \cite{Strom09, Teo09, Hou09, Liu11, Schmidt11, Dolcetto12, Dolcetto13, Chao13} together with induced spin-orbit coupling, \cite{Crepin12, Ilan12} strongly affects the QSH phase, leading to peculiar, and hopefully measurable, effects.
Among these, Qi \textit{et al.} \cite{Qi08} showed that when two barriers with antiparallel magnetizations are grown in contact with one helical edge state, a half-integer electron charge is trapped between these two, in contrast to the parallel magnetization case where the trapped charge is multiple of the electron charge.
This is a peculiar feature of the reduced degrees of freedom of HLLs. \cite{Meng09, Qi08}
A possible way to detect such half-charge is to employ Coulomb blockade measurements through a QSH quantum dot (QD): if the half-charge is present, a shift of the linear conductance peaks is expected. \cite{Qi08}\\
The system we consider in this paper is a QD realized between two magnetic barriers with different orientations of magnetization in HgTe QWs.
Due to the combined presence of helicity, magnetic barriers and quantum confinement, we demonstrate that oscillations appear in the in-plane spin density; these oscillations are washed out increasing the temperature, while they are enhanced by electron-electron interactions.
This is in contrast with traditional one-dimensional (1D) QDs, which show Friedel oscillations in the charge density, with integer charge. \cite{Fabrizio95} A possible way to detect Friedel oscillations in ordinary 1D QDs is by weakly coupling the dot to an atomic force microscope (AFM) tip. \cite{Halperin10, Boyd11, Traverso11, Mantelli12}
We propose to employ a magnetic force microscope (MFM) tip, \cite{Hubert97, Schwartz08} that is an AFM with a sharp magnetized tip, to probe such spin-density oscillations.
We show that by moving a magnetized tip, the shift in the chemical potential of the dot is sensitive to the spin density oscillations, thus allowing to probe them by transport measurements.
Furthermore, we show that the number of oscillations is related to the number of particles trapped inside the QD, thus allowing to distinguish the presence of the predicted fractional charge induced by the different magnetization. \cite{Qi08}\\
The paper is organized as follows. In Sec. \ref{model} we introduce the theoretical model of the QD. In Sec. \ref{density} we evaluate the expectation values of the spin densities, showing the appearance of oscillations in their in-plane components.
In Sec. \ref{MFMtip} we evaluate the correction to the chemical potential induced by the coupling with the magnetized tip.
Finally, Sec. \ref{conclusions} is devoted to the conclusions.

\section{Model}\label{model}
\begin{figure}[!ht]
\centering
\includegraphics[width=8cm,keepaspectratio]{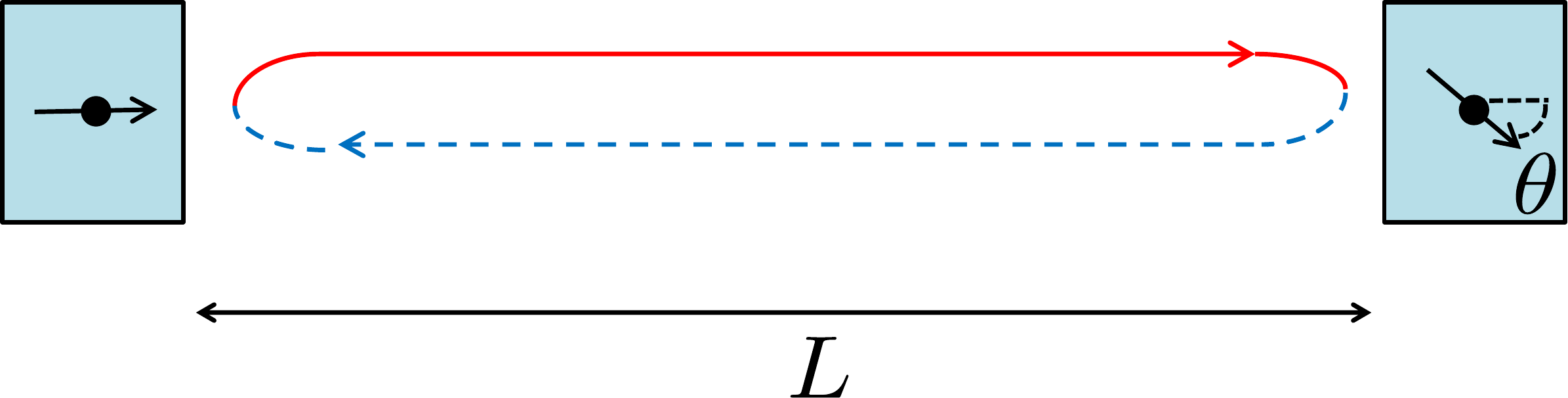}
\caption{(Color online) Schematic of the QSH dot. The magnetization of the left barrier points along the $x$-direction, while the magnetization of the right one forms an angle $\theta$ with the $x$ axis.}
\label{dot}
\end{figure}

We consider the helical edge state $\Psi=\left (\begin{matrix} \Psi_{R\uparrow} && \Psi_{L\downarrow} \end{matrix}\right )^{T}$, with right (left) moving spin up (down) electrons, described by the free Hamiltonian ($\hbar = 1$)
\begin{equation}
H_0=-i v_F\sigma_z\partial_x,
\end{equation}
where $v_F$ is the Fermi velocity and $\sigma_i$, with $i\in \{x,y,z\}$, are the Pauli matrices.
In order to create a QSH dot magnetic materials are required, \cite{Timm12} since the opening of a gap is related to TR symmetry breaking, which cannot occur with electrostatic gating.
The magnetic barriers are supposed to be narrower than the Fermi wavelength and localized at $x=0,L$, with Hamiltonian
\begin{equation}
H_{FM}=-m\left [ \delta(x)\sigma_x+\delta(x-L)\left ( \sigma_x\cos\theta-\sigma_y\sin\theta\right )\right ].
\end{equation}
This approximation holds if the magnetic barriers are narrower than few tens of nm, \cite{Timm12} which is nowadays an achievable size for magnetic nano-structures. \cite{Sattler11, Kaka05}
Here, we assume the magnetization of the barriers to lie in the $xy$ plane, with equal strength $m$ but pointing along different directions, as shown in Fig. \ref{dot}.
Note that the $z$-component of the spin is not a good quantum number, since it is not conserved in the presence of the magnetic barriers at the ends of the QSH dot. Indeed, these are responsible for backscattering, which, due to helicity, can only happen with spin flip.
In the limit ${m}/{v_F}\rightarrow\infty$ a QD is formed between the two magnetic barriers, and peculiar boundary conditions \cite{Timm12}
\begin{equation}\label{BCs}
\Psi_{L\downarrow}(0)=-i\Psi_{R\uparrow}(0), \ \ \ \
\Psi_{L\downarrow}(L)=ie^{-i\theta}\Psi_{R\uparrow}(L)
\end{equation}
arise.
The solutions of the Schr\"{o}dinger equation in the dot $H_0\Psi=E\Psi$, together with the conditions in Eq. \eqref{BCs}, are
\begin{equation}
\Psi_q(x)=\frac{1}{\sqrt{2L}}\left (\begin{matrix} e^{iqx} \\ -i e^{-iqx} \end{matrix}\right )
\end{equation}
with $\theta$-dependent momentum quantization
\begin{equation}
q=\frac{\pi}{L}\left (n-\frac{1}{2}+\frac{\theta}{2\pi}\right ) \ \ \ \ \ \ n\in\mathbb{Z}.
\end{equation}
The electronic (spinorial) field operator $\hat{\Psi}(x)$ can thus be expanded on the basis $\left \{\Psi_{q}(x)\right \}$ as $\hat{\Psi}(x)=\sum_{q}\Psi_q(x)\hat{c}_q$, the operator $\hat{c}_q$ destroying an electron with energy $v_F q$.
As a result of the backscattering induced by the magnetic barriers, spin up and spin down states are no longer independent; indeed one can introduce the operator \cite{Fabrizio95}
$\hat{\psi}(x)=\frac{1}{\sqrt{2L}}\sum_qe^{-iqx}\hat{c}_q$ and hence
\begin{equation}\label{Psi}
\hat{\Psi}(x)=\left (\begin{matrix}\hat{\psi}(-x) \\ -i\hat{\psi}(x)\end{matrix}\right ).
\end{equation}
This new field satisfies $\theta$-dependent boundary conditions over a double length $2L$
\begin{equation}\label{BC}
\hat{{\psi}}(L)=e^{i\pi\left (1-\frac{\theta}{\pi}\right )}\hat{{\psi}}(-L).
\end{equation}
Note that for parallel ($\theta=0$) and antiparallel ($\theta=\pi$) magnetizations, the field satisfies antiperiodic and periodic boundary conditions respectively.\\
Now we introduce the bosonized version \cite{VonDelft98} of $\hat{\psi}$.
By factoring out the oscillating phase factor $e^{-ik_Fx}$
\begin{equation}\label{psi_L}
\hat{\psi}(x)=e^{-ik_Fx}\hat{\psi}_L(x)
\end{equation}
with $k_F=\pi N_0/L$, $N_0$ being a reference number of particles in the dot, one has \cite{VonDelft98}
\begin{equation}\label{boso}
\hat{\psi}_L(x)=\frac{\hat{\mathcal{F}}}{\sqrt{2\pi\alpha}}e^{-i\pi\frac{x}{L}\left (\Delta \hat{N}-\frac{1}{2}+\frac{\theta}{2\pi}\right )}e^{-i\hat{\phi}(x)}
\end{equation}
with $\hat{\mathcal{F}}$ the Klein factor, $\Delta \hat{N}=\hat{N}-N_0$ the excess particle number with respect to $N_0$, and the $2L$-periodic bosonic field
\begin{equation}\label{phi}
\hat{\phi}(x)=-\sum_{k=\frac{\pi n}{L}>0}\sqrt{\frac{\pi}{kL}}e^{-\frac{\alpha}{2}k}\left [e^{-ikx}\hat{b}_k+e^{ikx}\hat{b}^{\dagger}_k\right ]
,\end{equation}
$\alpha= L/(\pi N)$ being a microscopic cut-off length and $\hat{b}_k$ the annihilation operators of the collective modes.
The particle density $
\hat{\rho}(x)=\hat{\Psi}^{\dagger}(x)\hat{\Psi}(x)$ is given by
\begin{align}\label{densL}
\notag\hat{\rho}(x)&=\hat{\psi}^{\dagger}_L(-x)\hat{\psi}_L(-x)+\hat{\psi}^{\dagger}_L(x)\hat{\psi}_L(x)\\
&=:\hat{\rho}(x):+\frac{N_{bg}}{L},
\end{align}
where the normal ordered term (denoted by $:\dots:$) takes into account the excess particle density with respect to the background $N_{bg}/L$,
\begin{equation}
:\hat{\rho}(x):=\frac{\Delta \hat{N}}{L}+\frac{1}{2\pi}\partial_x\left [\hat{\phi}(x)+\hat{\phi}(-x)\right ],
\end{equation}
\begin{equation}\label{bg}
N_{bg}=N_0+\frac{\theta}{2\pi}.
\end{equation}
Note that the background number $N_{bg}$ producing a fractional charge depends on the magnetic contribution $\theta/2\pi$. Indeed, for antiparallel magnetizations ($\theta=\pi$), half a background electron charge is between the barriers, in contrast to the parallel configuration ($\theta=0$), where the background charge between the barriers is a multiple of the electron charge: \cite{Qi08}
\begin{equation}
Q_{bg}=eN_{bg}=\left \{\begin{matrix}eN_0 && \theta=0 \\ eN_0+\frac{e}{2} && \theta=\pi \end{matrix}\right .
.\end{equation}
The free Hamiltonian of the dot can be recast as
\begin{eqnarray}
\hat{\mathcal{H}}_0&=&-i v_F\int_{0}^L dx :\hat{\Psi}^{\dagger}(x)\sigma_z\partial_x\hat{\Psi}(x):\nonumber \\
&=&v_F\sum_{k>0}k{\hat{b}}^{\dagger}_k{\hat{b}}_k+\frac{\pi v_F}{2L}\Delta {\hat{N}}\left (\Delta {\hat{N}}+\frac{\theta}{\pi}\right ).
\end{eqnarray}
It is now straightforward to include the presence of electron interactions
\begin{equation}\label{Hint}
\mathcal{\hat{H}}_i=\frac{g_{i}}{2}\int_0^L dx \left (\hat{\rho}(x)-\frac{N_0}{L}\right )^2,
\end{equation}
with $g_i$ proportional to the Coulomb repulsion.
The Hamiltonian $\mathcal{H} = \mathcal{H}_0+\mathcal{H}_i$ can be diagonalized through a Bogoliubov transformation of the collective modes: \cite{Giamarchi03}
\begin{equation}\label{H}
\mathcal{\hat{H}}=\mathcal{\hat{H}}_p+\mathcal{\hat{H}}_N,
\end{equation}
\begin{equation}\label{HpN}
\mathcal{\hat{H}}_p=\epsilon\sum_{k>0} m(k) \hat{a}^{\dagger}_k\hat{a}_k,\ \ \ \ \ \ \mathcal{\hat{H}}_N=\frac{E_N}{2}\left (\Delta \hat{N}+\frac{\theta}{2\pi}-n_g\right )^2
\end{equation}
with $\hat{a}_k$ the new plasmon operators, $m(k)=kL/\pi\in\mathbb{N}$.
The plasmon and the addition energies are $\epsilon=\pi v_F/(KL)$ and $E_N=\pi v_F/(K^2L)$ respectively, and scale differently by increasing the interaction, which is related to the Luttinger parameter through $K=\left [1+g_{i}/(\pi v_F)\right ]^{-1/2}$, with $K<1$ for repulsive interactions.
Here we also consider the presence of a gate voltage $V_{g}$, capacitively coupled to the dot, whose effect is to shift the background charge by $n_g\propto V_g$. \cite{Beenakker91, CavalierePRL04}
Note that the coupling to an external gate can lead to Rashba spin-orbit coupling, which may impact on the spin dynamics. \cite{Crepin12, Ilan12} However, this effect can be limited by the introduction of back gates, which allows to tune the Fermi level of the dot without inducing Rashba coupling. \cite{Vayrynen11}
The eigenstates $\left |\mathcal{S}\right \rangle=\left |N,\left \{n_k\right \}\right \rangle$ of the Hamiltonian Eq. \eqref{H} are identified by the excess particle number $\Delta \hat{N}=\hat{N}-N_0$ and by the occupation numbers $\left \{n_k\right \}$ of the collective modes.
The corresponding energies are
\begin{equation}\label{energy}
E\left (\mathcal{S}\right )=\frac{E_N}{2}\left (\Delta N+\frac{\theta}{2\pi}-n_g\right )^2+\epsilon\sum_{k>0}m(k)n_k
.\end{equation}
From Eq. \eqref{energy} one can extract the positions of the conductance peaks in the linear regime, where only transitions between ground states  $\left |N\right \rangle_{\mathrm{GS}}=\left |N,\left \{ n_k=0\right \}\right \rangle$ are relevant. The resonant condition $E(N+1)=E(N)$, with $E(N)\equiv E(N,\{ 0\})$, gives
\begin{equation}\label{ngres}
n_g^{(res)}=\Delta N+\frac{1}{2}\left (1+\frac{\theta}{\pi}\right )=\left \{\begin{matrix}\Delta N+\frac{1}{2} && \theta=0 \\ \Delta N+1 && \theta=\pi \end{matrix}\right .
.\end{equation}
As already suggested in Ref. \onlinecite{Qi08}, the presence of the half-charge can thus be revealed by observing shifted conductance peak positions as a function of $n_g$.

\section{Spin density oscillations}\label{density}
\begin{figure}[!ht]
\centering
\includegraphics[width=8.5cm,keepaspectratio]{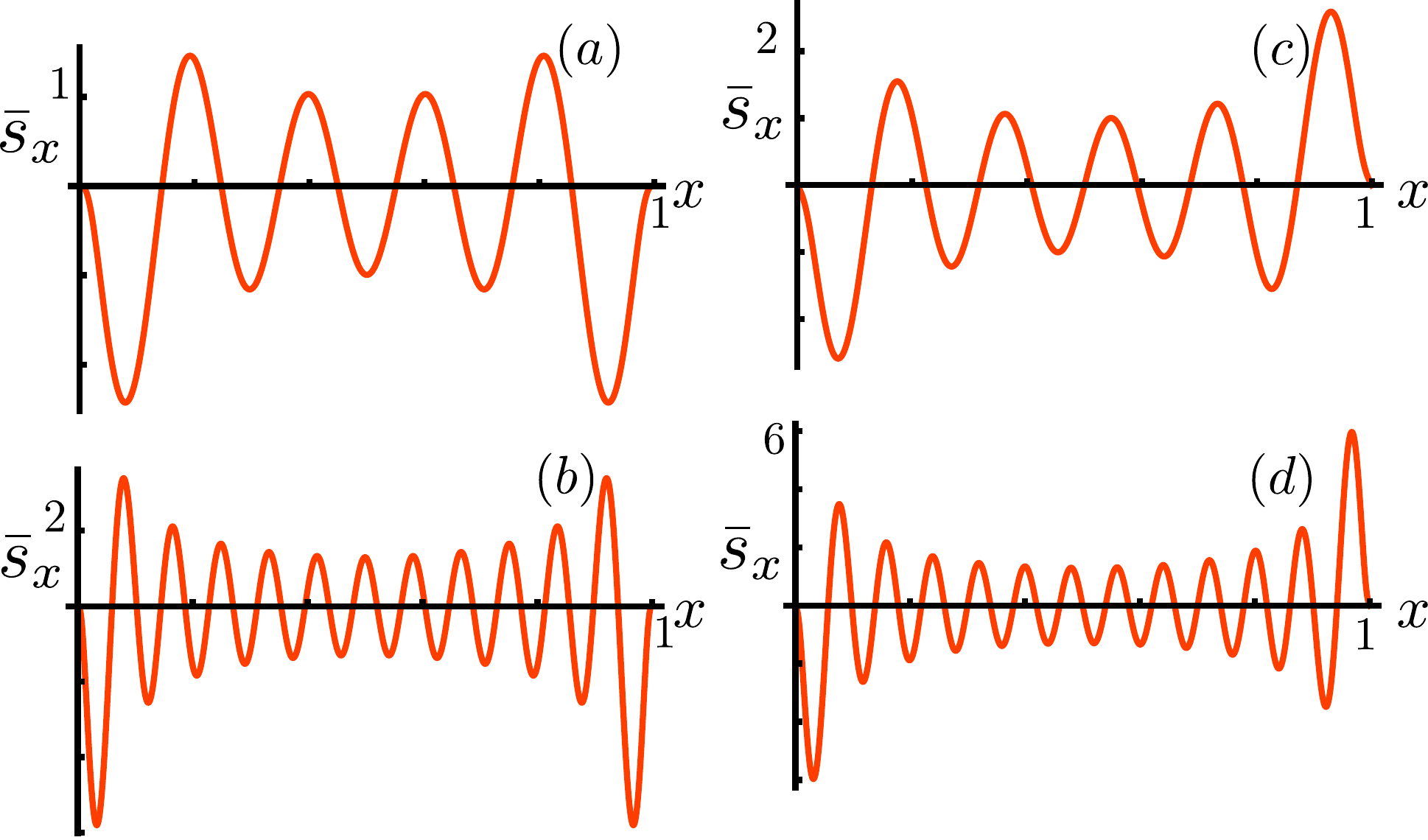}
\caption{(Color online) $\bar{s}_x(N,x)$ at $T=0$ in units
its $1/(2L)$ as a function of $x$ (units $L$) for $K=0.7$ and (a) $N=5$, $\theta=0$, (b) $N=12$, $\theta=0$, (c) $N=5$, $\theta=\pi$, (d) $N=12$, $\theta=\pi$.}
\label{sx_N_P}
\end{figure}

In this section we investigate the effects of quantum confinement on the spin density $\vec{\hat{s}}(x)=\hat{\Psi}^{\dagger}(x)\vec{\sigma}\hat{\Psi}(x)/2$.
We expect to observe peculiar features due to the combined presence of spin-momentum locking and magnetic barriers.
By recalling Eqs. \eqref{Psi} and \eqref{psi_L} one obtains
\begin{equation}
\hat{s}_x(x)=\frac{1}{2}\left (-ie^{-2ik_Fx}\hat{\psi}^{\dagger}_L(-x)\hat{\psi}_L(x)+H.c.\right ),
\end{equation}
\begin{equation}
\hat{s}_y(x)=\frac{1}{2}\left (-e^{-2ik_Fx}\hat{\psi}^{\dagger}_L(-x)\hat{\psi}_L(x)+H.c.\right ),
\end{equation}
\begin{equation}
\hat{s}_z(x)=\frac{1}{2}\left (\hat{\psi}^{\dagger}_L(-x)\hat{\psi}_L(-x)-\hat{\psi}^{\dagger}_L(x)\hat{\psi}_L(x)\right ).
\end{equation}
We are interested in the thermal expectation values of the above quantities at fixed $N$. Defining the density matrix $\hat{\rho}_{p} = \frac{1}{Z_{p}}e^{-\beta \mathcal{\hat{H}}_{p}}$, where $Z_{p} = \operatorname{tr}\left \{{e^{-\beta \mathcal{\hat{H}}_{p}}}\right \}$ is the partition function, the averages can be expressed as $\bar{s}_{i}(N,x) = \langle N|\operatorname{tr} \left \{ \hat{\rho}_{p}\hat{s}_{i}(x) \right \}|N\rangle$.
For temperatures $T$ such that $k_BT\ll \epsilon$, one has $\bar{s}_{i}(N,x) \approx \left\langle N \right|\hat{s}_{i}(x)\left| N \right\rangle_{\mathrm{GS}}$.\\
The $z$-component of the spin density is zero,  $\bar{s}_z(N,x)=0$.
On the other hand, oscillations appear in the in-plane ($xy$) spin density
\begin{eqnarray}\label{xspin}
\bar{s}_x(N,x)&=&-\sin\left [2\pi \frac{x}{L}\left (\Delta N+N_{bg}-\frac{1}{2}\right )-2f(x)\right ]\nonumber \\
&\times&\frac{1}{2\pi\alpha}e^{-\frac{1}{2}\left\langle \left [\phi(-x)-\phi(x)\right ]^2\right\rangle}
\end{eqnarray}
with $f(x)=\frac{1}{2}\tan^{-1}\left [\frac{\sin\left (2\pi x/L\right )}{e^{\pi\alpha/L}-\cos\left (2\pi x/L\right )}\right ]$.
The $y$-component $\bar{s}_y(N,x)$ is obtained from Eq. \eqref{xspin} with the replacement $\sin[\dots]\to\cos[\dots]$.
These oscillations emerge due to the presence of backscattering induced by the magnetic barriers, differently for example from Ref. \onlinecite{Soori12}, where in-plane spin density oscillations are predicted to occur in the presence of an external magnetic field.
At zero temperature the last term in Eq. \eqref{xspin} can be analytically evaluated
\begin{equation}\label{corr}
e^{-\frac{1}{2}\left\langle \left [\phi(-x)-\phi(x)\right ]^2\right\rangle}=\left (\frac{\sinh\left (\frac{\pi\alpha}{2L}\right )}{\sqrt{\sinh^2\left (\frac{\pi\alpha}{2L}\right )+\sin^2\left (\frac{\pi x}{L}\right )}}\right )^K.
\end{equation}
At finite temperature, we evaluate the bosonic correlator in Eq. \eqref{corr} numerically, while analytic evaluation could be addressed with the help of conformal field theory. \cite{Shankar90}\\
Figure \ref{sx_N_P} shows $\bar{s}_x$ at zero temperature for different particle numbers in the dot, for parallel and antiparallel barrier configurations.
\begin{figure}[!ht]
\centering
\includegraphics[width=6cm,keepaspectratio]{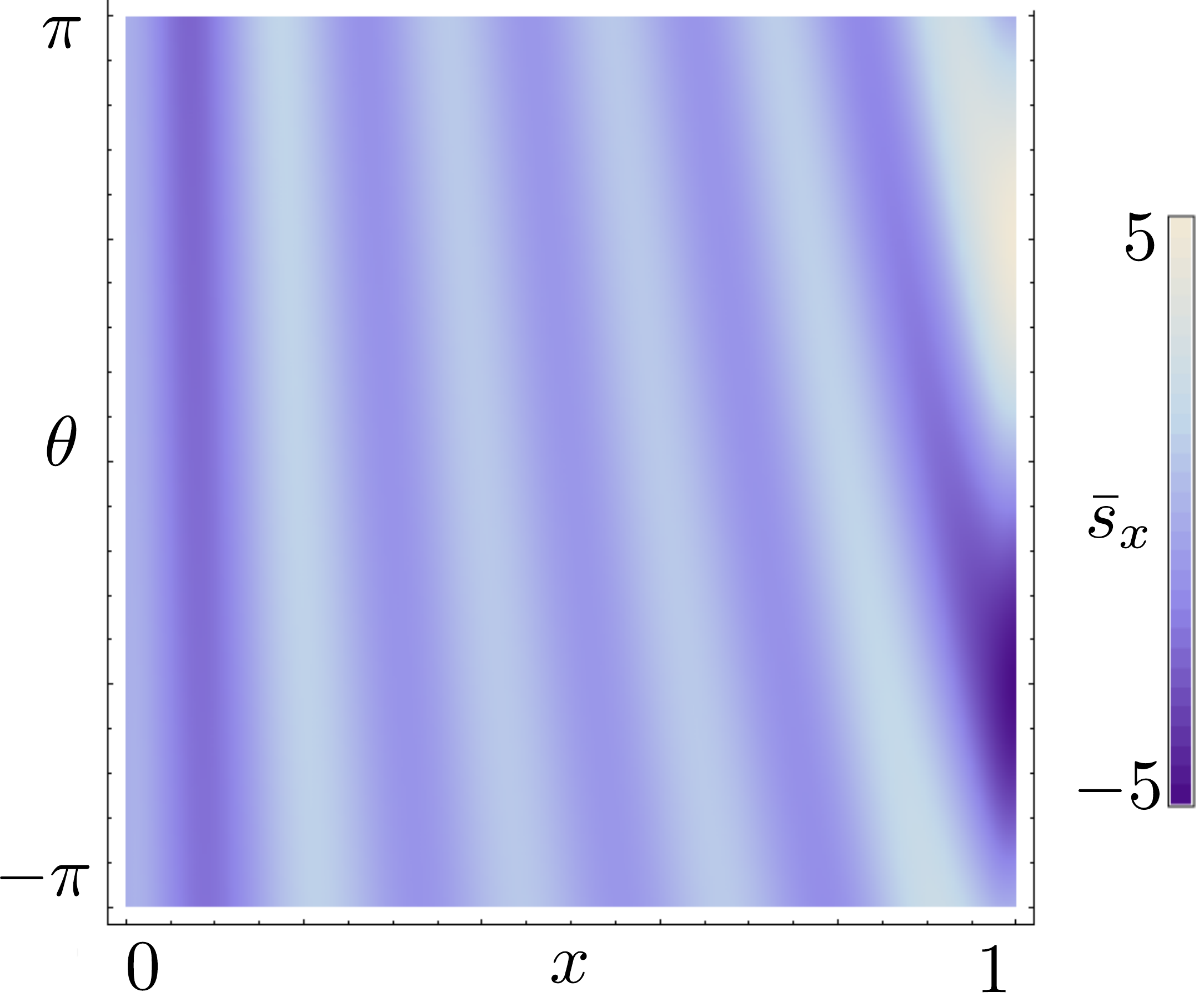}
\caption{(Color online) Density plot of $\bar{s}_x(5,x)$ at $T=0$ in units $1/(2L)$ as a function of $x$ (unit $L$) and $\theta$, for $K=0.7$.}
\label{sx_density}
\end{figure}
The number of oscillations is related to the number of particles inside the dot, as can be deduced by the first term in Eq. \eqref{xspin}.
It is worth to note that parallel and antiparallel configurations differ by half an oscillation: since the number of oscillations is related to the total charge in the dot (including the background), this reflects the fact that a fractional charge is trapped between the barriers in the antiparallel configuration, while an integer charge is trapped in the parallel configuration.\\
The evolution of the oscillations with the magnetization angle $\theta$ is displayed in Fig. \ref{sx_density}. By sweeping $\theta$ from $-\pi$ to $\pi$, the number of oscillations is increased by one: since the number of oscillations is related to the number of particles in the dot, all goes as if a unit charge gets trapped in the dot, as predicted. \cite{Qi08}\\
The interactions control the envelope of the oscillations through Eq. \eqref{corr}, but do not affect their number, since it only depends on the number of particles inside the dot, see Eq. \eqref{xspin}.
\begin{figure}[!ht]
\centering
\includegraphics[width=8.5cm,keepaspectratio]{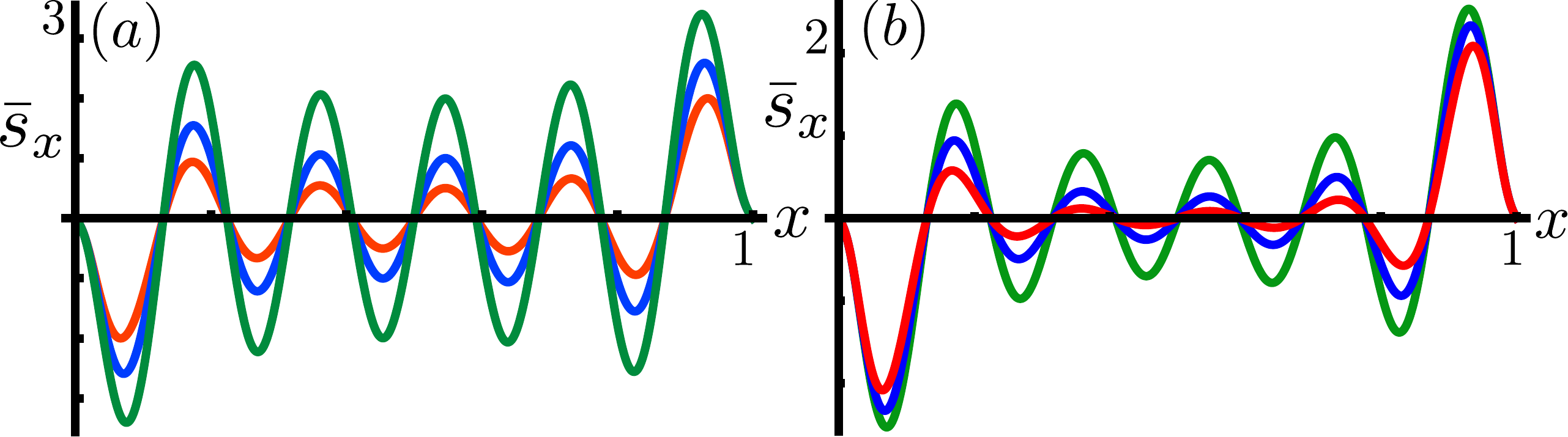}
\caption{(Color online) $\bar{s}_x(5,x)$ in units $1/(2L)$ as a function of $x$ (units $L$) in the antiparallel configuration $\theta=\pi$ for (a) $K=1$ (red), $K=0.7$ (blue) and $K=0.4$ (green), at fixed $T=0$; (b) $k_BT=\epsilon/2$ (green), $k_BT=\epsilon$ (blue), $k_BT=3\epsilon/2$ (red), at fixed $K=0.7$.}
\label{sx_K}
\end{figure}
Crucially, note that the oscillating behavior is enhanced by interactions, as shown in Fig. \ref{sx_K}(a), while it is damped by increasing the temperature, as shown in Fig. \ref{sx_K}(b).

\section{Probing the spin density oscillations with MFM tip}\label{MFMtip}
\begin{figure}[!ht]
\centering
\includegraphics[width=7cm,keepaspectratio]{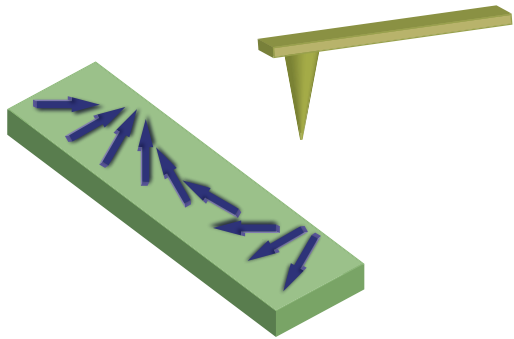}
\caption{(Color online) Schematic of the magnetized tip which couples with the magnetization of the dot.}
\label{schematip}
\end{figure}

\begin{figure}[!ht]
\centering
\includegraphics[width=8.5cm,keepaspectratio]{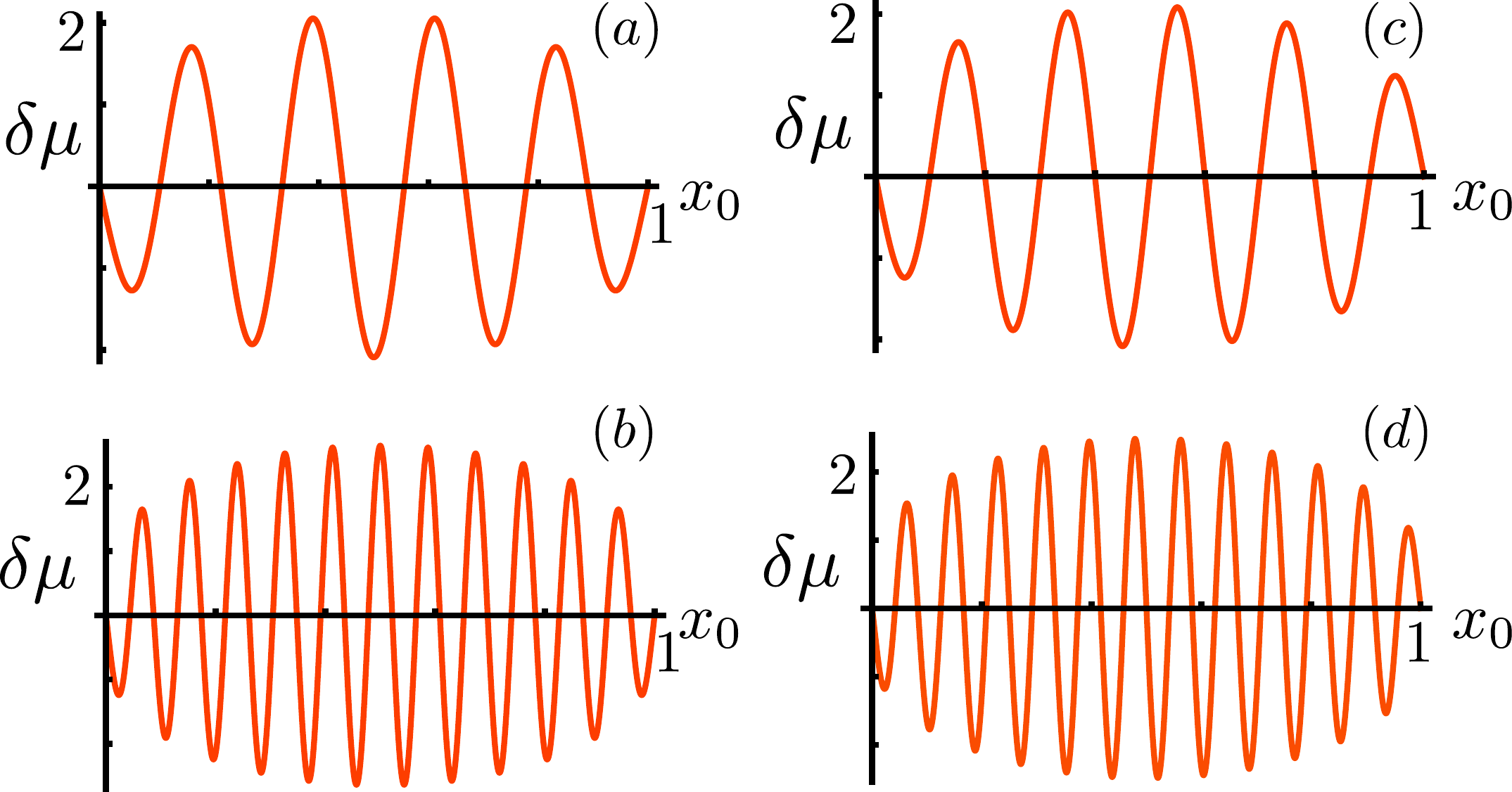}
\caption{(Color online) Chemical potential correction $\delta\mu$ at $T=0$ in units $- g\mu_B B/2$, as a function of the tip position $x_0$ (units $L$) for $K=0.7$, $\chi=0$ and (a) $N=4$, $\theta=0$, (b) $N=11$, $\theta=0$, (c) $N=4$, $\theta=\pi$, (d) $N=11$, $\theta=\pi$.}
\label{deltamu3D}
\end{figure}

We now show how a movable Magnetic Force Microscope (MFM) tip \cite{Hubert97, Schwartz08} could probe the presence of spin density oscillations in QSH dots.
It is well-known that an Atomic Force Microscope (AFM) tip can be exploited to gain information about the presence of charge oscillations in 1D dots. \cite{Halperin10, Boyd11, Traverso11, Mantelli12}
Indeed, by coupling with an AFM tip, the chemical potential, and consequently the position of the linear conductance peaks, become sensitive to the position of the tip.
In conventional 1D dots, the presence of Friedel oscillations is thus revealed by oscillations of the conductance peak positions as the tip moves.
In order to have direct indications about spin density oscillations, we propose to exploit a MFM tip capacitively coupled to the QSH dot: contrary to the AFM one, which couples to the particle density, the MFM tip consists in a magnetized tip which creates a localized magnetic field $\vec{B}_{tip}$ that couples with the magnetization of the dot, as schematically shown in Fig. \ref{schematip}.
We assume a tip narrower than the typical Fermi wavelength, with a magnetic field localized around its position. We then consider a local dot-tip coupling
\begin{equation}
\mathcal{\hat{H}}_{tip}(x_0)=-g\mu_BL\vec{B}_{tip}\cdot\vec{\hat{s}}(x_0),
\end{equation}
with $g$ the Landè $g$-factor of the HgTe QW, $\mu_B$ the Bohr magneton and $x_0$ the position of the tip along the dot.
The chemical potential of the QSH dot in the linear regime, for transitions with $N_{i(f)}$ particles in the initial (final) state, is defined as
\begin{equation}\label{mu}
\mu_{N_i\to N_f}(x_0)=E_{tot}\left (N_f,x_0\right )-E_{tot}\left (N_i,x_0\right )
\end{equation}
with $E_{tot}(N,x_0)$ the total energy in the presence of the tip.
Equation \eqref{mu} will be evaluated in the weak tip-coupling regime, namely $E_{tot}\left (N,x_0\right )=E\left (N\right )+\delta E\left (N,x_0\right )$, with $\delta E\left (N,x_0\right )=\left\langle N \right|\hat{\mathcal{H}}_{tip}(x_0)\left| N \right\rangle_{\mathrm{GS}}$ the lowest order correction (here considered at zero temperature, for simplicity).
The chemical potential is thus decomposed into the bare one $\mu^{(0)}_{N_i\to N_f}=E(N_f)-E(N_i)$ and the tip correction
\begin{equation}\label{deltamu}
\delta\mu_{N_i\to N_f}(x_0)=\delta E(N_f,x_0)-\delta E(N_i,x_0).
\end{equation}
This correction allows to probe spin density oscillations by moving the tip along the QSH dot, analogously to the AFM probe for Friedel oscillations in ordinary 1D dots.\\
Indeed consider, without loss of generality, \cite{note} the tip inducing the magnetic stray field $\vec{B}_{tip}=B\left (\cos\chi,\sin\chi,0 \right)$. The first order correction to the energy
\begin{eqnarray}\label{deltaEold}
&\delta E\left (N,x_0\right )=-g\mu_BL \left\langle N \right| \vec{B}_{tip}\cdot\vec{\hat{s}}(x_0) \left| N \right\rangle_{GS} \nonumber\\
&=-g\mu_BBL\left [\cos\chi \bar{s}_x(N,x_0)+\sin\chi \bar{s}_y(N,x_0)\right ]
\end{eqnarray}
depends on the expectation values of the in-plane components of the spin density, which have been evaluated and discussed in Sec. \ref{density}.
The correction in Eq. \eqref{deltaEold} can be easily rewritten as
\begin{eqnarray}\label{deltaE}
\delta E\left (N,x_0\right )&=&\sin\left [2\pi \frac{x_0}{L}\left (\Delta N+N_{bg}-\frac{1}{2}\right )-2f(x_0)+\chi\right ]\nonumber \\
&\times&\frac{g\mu_BBL}{2\pi\alpha}e^{-\frac{1}{2}\left\langle \left [\phi(-x_0)-\phi(x_0)\right ]^2\right\rangle}.
\end{eqnarray}
Then, from Eqs. \eqref{deltamu} and \eqref{deltaE}, we can evaluate the correction in the sequential regime $\delta\mu(N,x_0)\equiv\delta\mu_{N\to N+1}(x_0)$.\\
Figure \ref{deltamu3D} shows the correction $\delta\mu(N,x_0)$ as a function of the tip position for different number of particles in the dot, for parallel and antiparallel barrier configurations, when the tip couples to the $x$-component of the spin density only, i.e. $\chi=0$.
By virtue of the presence of the tip, the chemical potential correction shows oscillating behavior as the tip moves.
Note that the antiparallel configuration exhibits half an oscillation more that the parallel one. This feature is independent of the precise magnetization of the tip, since $\chi$ appears as a constant shift in Eq. \eqref{deltaE}, and is reminiscent of the absence or presence of the background half-charge.\\
Furthermore, by comparing with Fig. \ref{sx_N_P}, we note that $\delta\mu(N,x_0)$ closely follows the behavior of $\bar{s}_x(N+1,x_0)$. However, this holds only if $\chi\sim 0$, that is, when only $\hat{s}_x$ couples with the tip. In the opposite situation, when the tip couples with $\hat{s}_y$ only ($\chi\sim\pi/2$), the correction $\delta\mu(N,x_0)$ closely follows $\bar{s}_y(N+1,x_0)$.
Furthermore, just as happens for the spin density (see Fig. \ref{sx_K}a), the correction to the chemical potential due to the presence of the tip is enhanced by interactions.

\section{Conclusions}\label{conclusions}
We studied the properties of QSH dots, realized via two localized magnetic barriers as finite length edge states in two dimensional topological insulators.
The combined presence of magnetic barriers and spin-momentum locking, the hallmark of quantum spin Hall systems, leads to  an oscillating in-plane spin density. We showed that the number of oscillations depends on the particle number in the dot.
Parallel and antiparallel magnetization configurations differ by half an oscillation in the in-plane spin density, due to the presence of a half-charge trapped in the dot in the antiparallel configuration.
We proposed a method to detect these oscillations, by coupling the dot with a movable magnetic force microscope tip.
In the weak-coupling regime, we showed that the chemical potential of the dot has oscillatory corrections as the tip moves, directly connected to the spin density oscillations, which result in modified conductance peak position.
We showed that this effect is stable, and even enhanced, in the presence of interactions.

\section*{Acknowledgments}
We thank D. Ferraro and A. Braggio for useful discussions. The support of CNR STM 2010 program, EU-FP7 via Grant
No. ITN-2008-234970 NANOCTM, CNR-SPIN via Seed Project PGESE001 and MIUR-FIRB - Futuro in Ricerca 2012 - Project HybridNanoDev (Grant  No.RBFR1236VV) is acknowledged.


\begin{thebibliography}{10}
\bibitem{note} An out-of-plane component $B_z$ does not contribuite to Eq. \eqref{deltaEold}, since it gives a term proportional to $\bar{s}_z$, which is zero.
\end{thebibliography}
\end{document}